# Visible Brillouin-quadratic microlaser in a high-Q thin-film lithium niobate microdisk

Xiaochao Luo,[1,5,11] Chuntao Li,[2,4,11] Xingzhao Huang,[3] Jintian Lin,[1,5,*] Renhong Gao,[2] Yifei Yao,[6] Yingnuo Qiu,[1,5] Yixuan Yang,[1,5] Lei Wang,[6] Huakang Yu,[3,‡] And Ya Cheng[1,2,4,7,8,9,10,†]

[1]*State Key Laboratory of Ultra-intense Laser Science and Technology and CAS Center for Excellence in Ultra-Intense Laser Science, Shanghai Institute of Optics and Fine Mechanics (SIOM), Chinese Academy of Sciences (CAS), Shanghai 201800, China*

[2]*The Extreme Optoelectromechanics Laboratory (XXL), School of Physics and Electronic Science, East China Normal University, Shanghai 200241, China*

[3]*School of Physics and Optoelectronics, State Key Laboratory of Luminescent Materials and Devices, South China University of Technology, Guangzhou 510460, People's Republic of China*

[4]*State Key Laboratory of Precision Spectroscopy, East China Normal University, Shanghai 200062, China*

[5]*Center of Materials Science and Optoelectronics Engineering, University of Chinese Academy of Sciences, Beijing 100049, China*

[6]*School of Physics, State Key Laboratory of Crystal Materials, Shandong University, Jinan 250100, China*

[7]*Shanghai Research Center for Quantum Sciences, Shanghai 201315, China*

[8]*Hefei National Laboratory, Hefei 230088, China*

[9]*Collaborative Innovation Center of Extreme Optics, Shanxi University, Taiyuan 030006, China*

[10]*Collaborative Innovation Center of Light Manipulations and Applications, Shandong Normal University, Jinan 250358, China*

[11]*Xiaochao Luo and Chuntao Li contributed equally to this work.*

*\*E-mail: jintianlin@siom.ac.cn*

*‡E-mail: hkyu@scut.edu.cn*

*†E-mail: ya.cheng@siom.ac.cn*

**Narrow-linewidth lasers at short/visible wavelengths are crucial for quantum and atomic applications, such as atomic clocks, quantum computing, atomic and molecular spectroscopy, and quantum sensing. However, such lasers are often only accessible in bulky tabletop systems and remain scarce in integrated photonic platform. Here, we report an on-chip visible Brillouin-quadratic microlaser in a 117-μm-diameter thin-film lithium niobate (TFLN) microdisk via dispersion engineering. Enabled by the ultra-high Q factor of $4.0 \times 10^6$ and small mode volume, strong photon-phonon interaction and high second-order nonlinearity of the TFLN microdisk, narrow-linewidth Stokes Brillouin lasing (SBL) is demonstrated with 10.17 GHz Brillouin shift under a 1560-nm pump, exhibiting a short-term narrow linewidth of 254 Hz and a low threshold of only 1.81 mW. Meanwhile, efficient second harmonic generation (SHG) of the SBL signal is also observed at 780 nm, with a normalized conversion efficiency of 3.61%/mW, made possible by simultaneous phase matching fulfillments for both narrow-linewidth SBL and its SHG. This demonstration of an integrated ultra-narrow linewidth visible wavelength Brillouin-quadratic lasers opens new avenues toward chip-scale quantum information processing and precise metrology.**

Narrow linewidth laser sources operating at short/visible wavelengths are highly desirable for achieving the high spectral purity required in precision atomic, molecular and optical (AMO) physics and associated applications[1-7], including atomic clocks, quantum computing, atomic and molecular spectroscopy, and quantum sensing. Traditionally, such narrow-linewidth lasers have been realized using table-top systems stabilized to vapor cells or large optical reference cavities[8-11], achieving the ultrahigh spectral purity necessary for probing narrow optical clock transitions in cold atoms. However, as AMO experiments grow in complexity and as efforts intensify toward portable or even autonomous optical clocks, these bulky laboratory-scale systems face critical limitation in size, robustness and scalability. There is a growing demand for compact, reliable short/visible wavelength lasers capable of supporting increasing experimental complexity, including multiple operating wavelengths and larger amount of atoms or molecules involved. Photonic integration offers a promising solution by enabling chip-scale of laser systems while improving reliability and reducing sensitivity to environmental disturbances within a chip-scale footprint and with low power consumption[12-14].

Integrated stimulated Brillouin scattering (SBS) lasers leverage the narrow-bandwidth Brillouin gain of sub-100 megahertz and the ultra-high Q factor of the on-chip resonators to realize ultra-low phase noise performance[15-25]. These integrated stimulated Brillouin lasers have been predominantly demonstrated in the telecom band[15-25], where they exhibit ultra-narrow fundamental linewidth lasing and support stable Kerr-Brillouin comb formation. It is noted that the operating wavelengths of stimulated Brillouin lasers could be expanded to visible and mid-infrared spectral ranges[26,27]. To date, bi-chromatic light emissions spanning both the telecom and short/visible bands are still out of reach in a single integrated Brillouin laser. Key bottlenecks include the lack of ultra-high Q resonators in the short/visible band, and the difficulties in simultaneously satisfying phase match conditions for both Brillouin lasing[24] and subsequent quadratic nonlinear interactions[28-30].

In this work, we demonstrate bi-chromatic Brillouin microlaser operating simultaneously in the telecom and visible bands, realized in an on-chip thin-film lithium niobate (TFLN) microdisk. This Brillouin-quadratic microlaser demonstration relies on advances in precise dispersion engineering and low-loss fabrication of $\chi^{(2)}$ microresonator with small mode volume. The suspended TFLN microdisk, with a compact diameter of 117 μm, is dispersion engineered to enable phase-matched Stokes Brillouin lasing (SBL) via strong optomechanical coupling with two orthogonally polarized fundamental cavity modes, while simultaneously generating the second harmonics of the SBL via modal phase match. The designed TFLN microdisk is fabricated with ultra-smooth surface by photolithography assisted chemo-mechanical etching method[31], featuring ultrahigh Q factors of $4.0\times10^6$ in the telecom band and $1.3\times10^6$ at its second harmonic, facilitating significant linewidth narrowing, reduced lasing threshold and resonant SHG. Consequently, SBL at 1559.718 nm is demonstrated with a 1.81 mW optical threshold, and a short-term linewidth of 254.365 Hz, under 1559.632 nm optical pump. Meanwhile, SHG of this SBL signal was observed at 779.859 nm, with a normalized conversion efficiency of 3.61%/mW. This Brillouin-quadratic microlaser demonstration paves the way toward compact and multifunctional quantum and atomic photonic systems.

The on-chip TFLN microdisk was fabricated on a Z-cut thin-film lithium niobate on insulator (LNOI) using femtosecond laser photolithographic-assisted chemo-mechanical etching (PLACE) technique[31]. Further details are provided in the **Methods** section. An optical microscope image of the fabricated TFLN microdisk supported by a small silicon dioxide pillar to suppress the leakage of the acoustic modes to the substrate[32-34], is depicted in the inset of Fig. 1. The TFLN microdisk has a diameter of ~117 μm and a thickness of ~800 nm, dispersion engineered to simultaneously fulfill phase matching and energy conversation conditions for both the backward SBL[24] and its SHG processes (i.e., Brillouin-quadratic microlaser formation), thereby enabling Brillouin-quadratic microlaser operation.

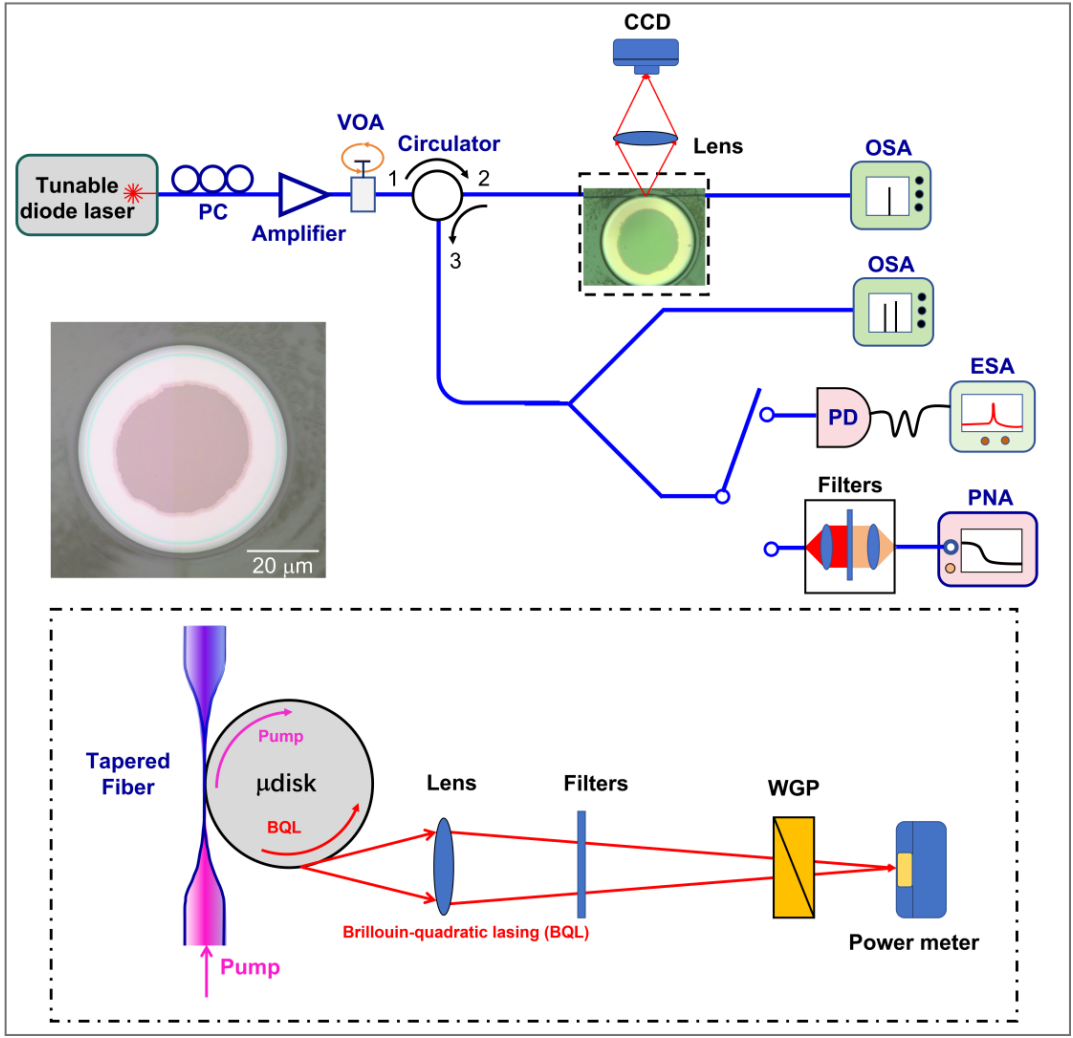

**FIG. 1: Schematic of the setup for the generation of Brillouin-quadratic microlaser in the microdisk.** PC, polarization controller; VOA, variable optical attenuator; CCD, charge coupled device; OSA, optical spectrum analyzer; PD, photodetector; ESA, electrical spectrum analyzer; PNA, phase noise analyzer. Inset (middle left): Optical microscope image of the fabricated microdisk. Inset (bottom): Power measurement and polarization analysis of the backward Brillouin-quadratic lasing (BQL) signal scattering from the edge of the microdisk. WGP, wire grating polarizer.

The experimental setup for the formation of Brillouin-quadratic microlaser in the suspended LNOI microdisk resonator is illustrated in Fig. 1. A continuous wave (CW) tunable laser (New Focus Inc., Model TLB6728) in the telecom band was used as the pump laser source to excite the cascaded nonlinear processes in the microdisk resonator. Before the light from the CW laser was launched in the optical circulator, it propagates through an in-line fiber polarization controller (FPC562, Thorlabs Inc.) and an erbium-doped fiber amplifier (EDFA), followed by a variable optical attenuator (VOA) to modify the polarization, and achieve variable pump power. The pump light was then sent into the microdisk resonator via port 2 of the circulator using an optical tapered fiber with a waist of 2 um. It should be emphasized that the tapered fiber was carefully placed on the edge of the microdisk resonator and was in direct contact with the resonator to couple light in and out of the resonator. The coupling position of the tapered fiber is very crucial for achieving cross-polarized modal phase match for Brillouin-quadratic processes[24]. The generated backward SBL signal was collected by the circulator, and then detected by an optical spectrum analyzer (AQ6370D, Yokogawa Inc.) with 0.02-nm resolution for spectral analysis. And a high-speed photodetector connected with a real-time electrical spectrum analyzer (Tektronix RAS5115B) was used to measure the beat note microwave signal by beating the pump light and the SBL. The linewidth of the backward SBL was further measured with a commercially available laser phase noise analyzer (PNA) using tunable optical filters to block the pump light. The details for the measurement of Brillouin-quadratic lasing signal, including spectrum analysis, polarization analysis, and power measurement, can be found in **Methods** section.

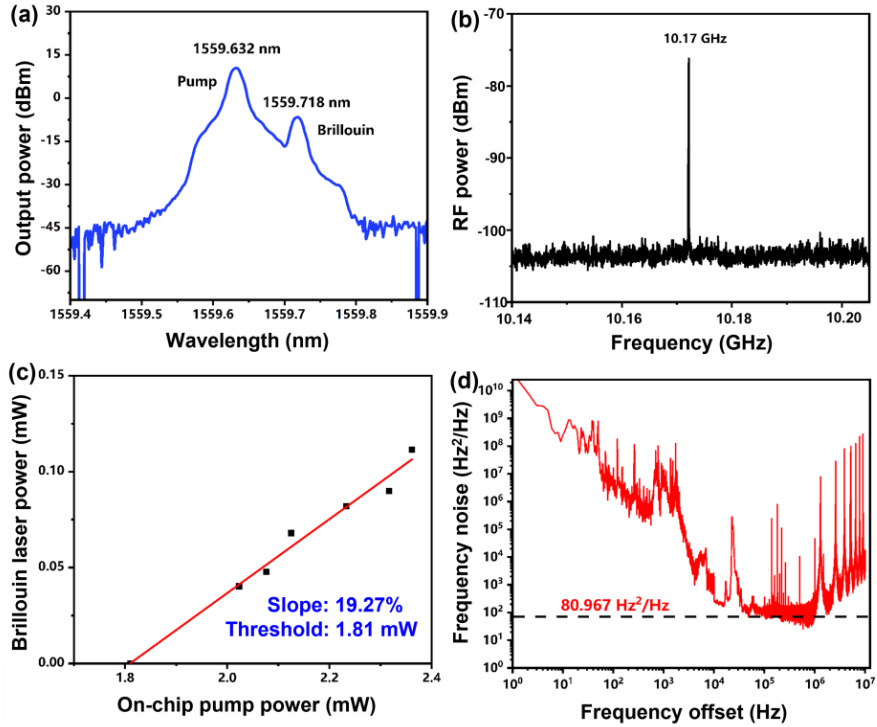

**FIG 2: Backward Stokes Brillouin lasing (SBL). a**, Optical spectrum of the SBL signal. **b**, Microwave beat signals of the pump wave and the SBL signal. **c**, Output power of SBL vs. on-chip pump power. **d**, The frequency noise spectrum of the backward SBL.

When the pump light injected into the microdisk resonator was tuned to 1559.632 nm and the pump power exceeded the threshold of SBL, two distinct nonlinear phenomena were concurrently observed, including a backward-propagating SBL signal and its SHG signal. The backward-propagating SBL signal was detected at 1559.718 nm, as shown in Fig. 2(a). The wavelength interval between the pump light and the SBL signal was approximately 0.08 nm, corresponding to a Stokes Brillouin shift $\Omega_B$ of ~10 GHz. And this Brillouin shift $\Omega_B$ was further confirmed by the detected radio-frequency (RF) beat note microwave signal with finer resolution by means of optical heterodyne method using the electrical spectrum analyzer (ESA). This RF signal centered at 10.17-GHz frequency, as shown in Fig. 2(b). Moreover, the output power of the backward SBL signal varied with different pump powers was also recorded to characterize the threshold behavior of the backward SBL, as plotted in Fig. 2(c). When the pump power exceeds the threshold, the output power of the backward SBL (black dots) increases linearly with the pump power. Through linear fitting (red line), the threshold of the backward SBL is determined to be 1.81 mW, and the conversion efficiency reaches 19.27%.

Furthermore, the frequency noise of the backward SBL was measured using a correlated self-heterodyne method with $\beta$-separation line[22], as depicted in Fig. 2(d). The white-frequency-noise floors $N_{wfn}$ of the backward SBL was quantified as 81 Hz$^2$/Hz, revealing short-term linewidths $L_{st}$ of 254 Hz ($L_{st} = N_{wfn} \times \pi$). While the short-term linewidth of the pump laser was measured as ~ 500 Hz, the obtained SBL exhibited a clear linewidth narrowing, which is attributed to the narrow Brillouin gain bandwidth of TFLN, ultrahigh optical quality factors of the microcavity and the long lifetime of the Brillouin phonon[15-17].

Simultaneously, only one shortwave signal was observed at 779.859 nm with the high spectral resolution 0.02 nm of the OSA, as shown in Fig. 3(a). This wavelength is precisely equal to second harmonics of the SBL signal. As a result, this signal was ascribed to the SHG of the SBL signal. The side-view optical microscope image of the microdisk is depicted in the inset of Fig. 3(b), showing obvious visible light emission leaked from each edge of the microdisk. These light spots exhibit spatial high-order mode characteristics. Moreover, the light spots at the right edge are much brighter than the left ones, revealing that the clockwise signal is much stronger than the counter-clockwise one. The counter-clockwise SHG signal was believed to be presented due to cavity-enhanced Rayleigh scattering in the ultrahigh-Q microdisk[35,36]. We recorded the output power of the SHG signal by collecting the scattered light from the right edge of the microdisk at different pump powers, so as to determine power dependence behavior. Figure 3(b) shows a quadratic dependence of the SHG signal on the SBL power. The measured SHG conversion efficiency (black dots) increases linearly with the SBL power, as shown in Fig. 3(b). By linear fitting (red line), the normalized SHG conversion efficiency is determined to be 3.61%/mW. The maximum output power of the SHG signal reached an output power of 2.83 μW with an absolute conversion efficiency of 1.01% when the SBL power was 0.28 mW (corresponding to an on-chip pump level of 3.028 mW). It is worth noting that no SHG signal of the pump light was detected, which was attributed to the absence of phase matching and the difficulty in satisfying the dual-resonance condition.

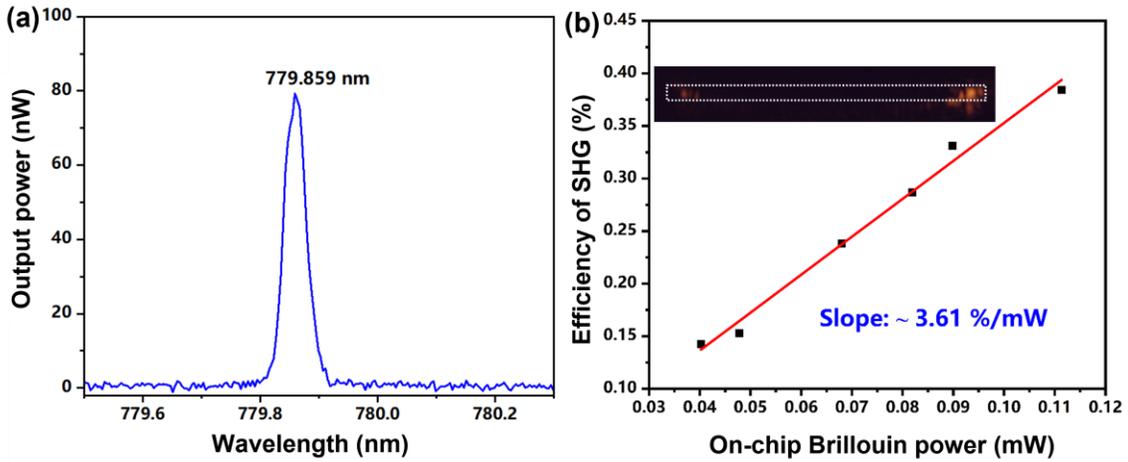

**FIG. 3: Visible Brillouin-quadratic microlaser. a**, Spectrum of the forward SHG of the Brillouin lasing signal by Rayleigh backscattering. **b**, Conversion efficiency of SHG as a function of the on-chip SBL power. Inset: Side-view optical microscope image of the scattered light spots from the edge of the microdisk, showing high-order spatial modes, with the dashed box denotes the profile of the microdisk (the vertical dimension of the box is enlarged with 6.8 times).

To evaluate the mode structures of the microdisk, two continuously tunable lasers around 1560 nm and 780 nm wavelengths were used for exciting the involved modes by scanning the wavelength cross the modes participated in the visible Brillouin-quadratic lasing, respectively. The pump power injected into the microdisk resonator was set to as low as 5 μW, the transmission spectra of the microdisk resonator were obtained, as shown in Fig. 4. The transverse-electric (TE) pump mode and the transverse-magnetic (TM) SBL mode are depicted in Fig. 4(a), showing a wavelength of 0.086 nm, agreeing well with aforementioned results. The Lorenz fitting (red curves) shows that the loaded Q factors of the pump mode, SBL mode were determined to be $4.00\times10^6$ and $3.27\times10^6$, respectively. Meanwhile, the mode structure around the SHG mode of the Brillouin-quadratic lasing signal is shown in Fig. 4(d). There is actually a high-order mode resonant with SHG of the Brillouin-quadratic lasing signal at 779.859 nm, and the loaded Q factor was measured as high as $1.35×10^6$. The ultrahigh Q factors guarantee a considerable increase in the intracavity built-up pump power within the small microdisk and boost high conversion efficiencies, which are crucial for low-threshold SBL and efficient Brillouin-quadratic lasing. It is worth noting that there is no mode resonant with the SHG of the pump light or the sum-frequency generation of the pump light and the Brillouin lasing signal.

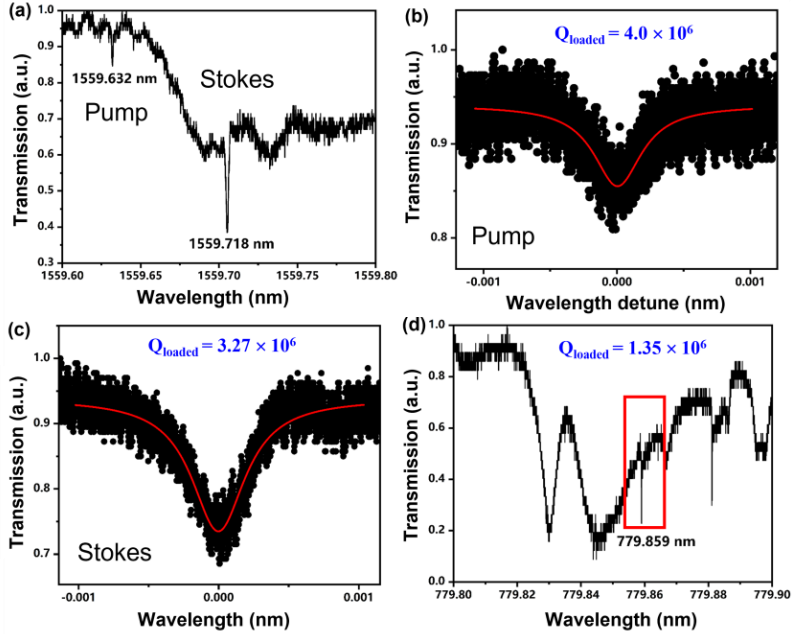

**FIG. 4: Mode structures of the microdisk. a**, Transmission spectrum of the pump mode and the SBL mode. **b**, Loaded Q factors of the pump mode. **c**, Loaded Q factor of the SBL mode. **d**, Transmission spectrum around 779.85 nm, with the red box denotes the SHG mode of the SBL signal, and there is no mode resonant with the SHG of the pump light.

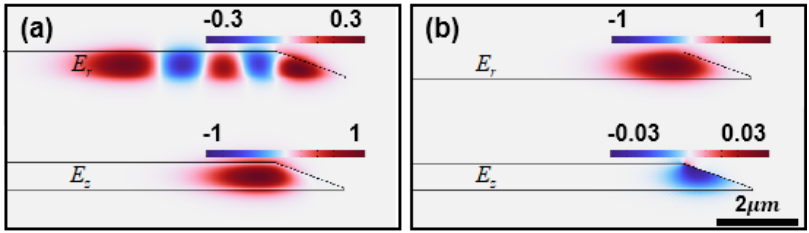

**FIG. 5: Electric field distribution of the optical modes**. Electric field distribution of the optical modes for (a) the pump light ($TM_0$-polarized, m = 421 @ 1559.632 nm) and (b) the SBL modes ($TE_0$-polarized, m = -454 @ 1559.718 nm). The distributions of the electric field components, $E_r$ and $E_z$, are shown for both the pump light and the SBL modes.

It is necessary to reveal phase match schemes for the Brillouin-quadratic lasing process. Here, cross-polarized stimulate Brillouin scattering process involved with the fundamental TM optical mode ($TM_{0,421}$ as pump mode), fundamental TE optical mode ($TE_{0,-454}$ as SBL mode) was leveraged in the small microdisk and shear mechanical mode, ensuring strong optomechanical coupling rate[24]. The mode distributions of the pump light optical mode and the backward-propagating SBL mode were obtained through numerical simulations, as shown in Fig. 5. Moreover, we also performed numerical simulations showing the optical field distribution of the Brillouin-quadratic lasing mode, which is determined to $TM_{4,-908}$ mode, as illustrated in Fig. 6. This cross-polarized configuration of SHG process could utilize the second-order nonlinear coefficient of $d_{31}$. And the SHG process of the pump light, the sum-frequency generation of the pump light and the Brillouin lasing signal are prohibited due to the absence of dual-resonance rules.

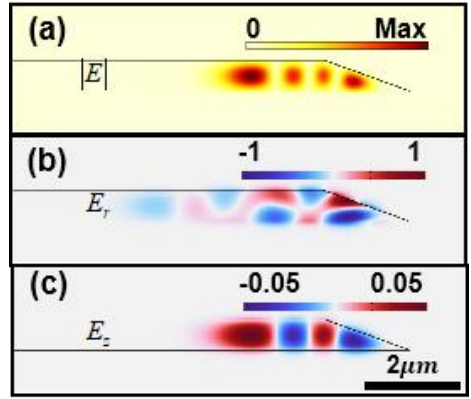

**FIG 6: Optical mode profile distribution of the Brillouin-quadratic lasing mode** ($TM_4$-polarized, $m = -908$ @ 779.859 nm). Normalized electric field components (a) $E$, (b) $E_r$ and (c) $E_z$ of the optical mode.

In summary, benefiting from the strong $\chi^{(2)}$ nonlinearity and photon-phonon interaction along with careful dispersion engineering, Brillouin-quadratic microlaser is demonstrated in the high-Q LNOI microdisk resonator. The backward SBL is generated with a record-low threshold of only 1.81 mW on the TFLN platform so far[24,25], accompanied by efficient SHG with a normalized conversion efficiency of 3.61%/mW. By demonstrating the coexistence of Brillouin lasing and SHG within a single high-Q LNOI microcavity, we establish a new paradigm for the design of efficient and compact nonlinear photonic integrated systems. This work paves the way toward advanced photonic devices integrating diverse functionalities, such as narrow-linewidth lasers, frequency converters, and quantum light sources, all within a single photonic chip.

## Methods

**Fabrication.** The microdisk is fabricated by photolithography assisted chemo-mechanical etching (PLACE). The fabrication process begins from the preparation of Erbium ion doped lithium niobate (LN) thin film wafer by ion slicing. And the sample endures chromium (Cr) layer coating, hard mask patterning via femtosecond laser ablation, pattern transferring from the Cr hard mask to LN thin film via chemo-mechanical polishing, and chemical wet etching to completely remove the Cr layer and partially remove the silica layer underneath the LN disk into the supporting pedestal. The details of the fabrication can be found in Ref. [31].

**Measurement.** Since the limited detectable bandwidth of the circulator, the SHG of the backward SBL does not propagate through the circulator. Consequently, it is difficult to detect the spectrum of the backward propagating SHG directly. However, because of strong Rayleigh backscattering of the backward SHG signal circulating within the high-Q microdisk[35,36], there was a weak forward propagating SHG signal, and was detected by the OSA. Meanwhile, an objective lens of numerical number of 0.25 was used to collect the scattered SHG from the edge of the microdisk for power measurement. And optical filters and an optical polarizer were inserted before a power meter for blocking the infrared signals and polarization analysis, respectively.

(2017).

7. Fan, H. et al. Atom based RF electric field sensing. *J. Phys. B*. **48**, 202001 (2015).

8. Fox, R. W., Oates, C. W. & Hollberg, L. W. Stabilizing diode lasers to highfinesse cavities. in *Experimental Methods in the Physical Sciences* (eds. van Zee, R. D. & Looney, J. P.) vol. 40 1–46 (Academic Press, 2003).

9. Ludlow, A. D. et al. Compact, thermal-noise-limited optical cavity for diode laser stabilization at $1\times10^{-15}$. *Opt. Lett.* **32**, 641–643 (2007).

10. Yao, Y., Jiang, Y., Yu, H., Bi, Z. & Ma. L. Optical frequency divider with division uncertainty at the $10^{-21}$ level. *Nat. Sci. Rev.* **3**, 463–469 (2016).

11. Nicholson, T. L. et al. Systematic evaluation of an atomic clock at $2 \times 10^{-18}$ total uncertainty. *Nat. Commun.* **6**, 6896 (2015).

12. Li, M. et al. Integrated Pockels laser. *Nat. Commun.* **13**, 5344 (2022).

13. Han, Y., Park, H., Bowers J. & Lau, K. M. Recent advances in light sources on silicon. *Adv. Opt. Photon.* **14**, 404–454 (2022).

14. Blumenthal, D. J. Photonic integration for UV to IR applications. *APL Photon.* **5**, 020903 (2020).

15. Gundavarapu, S. et al. Sub-hertz fundamental linewidth photonic integrated Brillouin laser. *Nat. Photonics* **13**, 60 (2019).

16. Yang, K. Y. et al. Bridging ultrahigh-Q devices and photonic circuits. *Nat. Photonics* **12**, 297–302 (2018).

17. Grudinin, I. S., Matsko, A. B. & Maleki, L. Brillouin Lasing with a $CaF_2$ whispering gallery mode resonator. *Phys. Rev. Lett.* **102**, 043902 (2009).

18. Zhu, S., Xiao, B., Jiang, B., Shi, L. & Zhang X. Tunable Brillouin and Raman microlasers using hybrid microbottle resonators. *Nanophotonics* **8**, 931–940 (2019).

19. Otterstrom, N. T., Behunin, R. O., Kittlaus, E. A., Wang, Z. & Rakich, P. T. A silicon Brillouin laser. *Science* **360**, 1113–1116 (2018).

20. Bai, Y. et al. Brillouin-Kerr soliton frequency combs in an optical microresonator. *Phys. Rev. Lett.* **126**, 063901 (2021).11

# Acknowledgments

We thank Peking university Yangtze delta institute of optoelectronics for providing the laser phase noise measurement system (iFN5000).